\documentclass[11pt]{article}
\usepackage{graphicx} 
\usepackage{amsmath, amssymb}
\usepackage{authblk}
\usepackage{mathtools}
\usepackage{physics}
\usepackage{hyperref}
\usepackage{cite}
\usepackage{geometry}
\usepackage{cleveref}
\hypersetup{
  colorlinks=true,
  linkcolor=blue,
  citecolor=blue,
  urlcolor=blue
}

\title{Hilbert Series and Complete-Intersection Structure of Coulomb Branches for Non-Maximal Nilpotent Orbits of $SL(N)$}

\author{Ayush Kumar}
\affil{Department of Physics and Astronomy, National Institute of Technology, Rourkela - 769008, India}

\date{}

\begin{document}

\maketitle

\begin{abstract}
We study the Coulomb branches of three-dimensional $\mathcal N=4$ quiver gauge
theories of type $T_\rho(SU(N))$ associated with non-maximal nilpotent orbits of
$SL(N)$. Using the Hall--Littlewood closed form for Coulomb-branch Hilbert series,
together with independent checks from the monopole formula, we compute exact
unrefined Hilbert series for all non-maximal partitions $\rho\vdash N$ with $N=4$,
and extend the analysis to $N=5,6$. 

By analyzing the plethystic logarithms of the resulting Hilbert series, we find
that in all cases examined the Coulomb branch is a complete intersection. The
number of generators and relations follows a uniform pattern governed by the
transpose partition $\rho^T$, with exactly $N-1$ relations appearing
independently of $\rho$ in these examples. We summarize the results in explicit
classification tables and formulate conjectures extending these patterns to
arbitrary $N$. Our findings provide strong evidence for a remarkable uniformity in
the algebraic structure of Coulomb branches within the $T_\rho(SU(N))$ family at
low rank.
\end{abstract}

\section{Introduction}

The moduli spaces of vacua of three-dimensional $\mathcal N=4$ supersymmetric gauge theories exhibit a rich geometric structure and have attracted sustained attention in recent years. In particular, the Coulomb branch of such a theory is a hyperk\"ahler cone which receives quantum corrections and does not admit a direct classical description in terms of the Lagrangian fields. A rigorous mathematical framework for Coulomb branches was developed by Braverman, Finkelberg, and Nakajima (BFN), who constructed Coulomb branches as affine algebraic varieties associated with pairs $(G,\mathcal R)$, where $G$ is a complex reductive group and $\mathcal R$ is a symplectic representation \cite{BFN1,BFN2}. This construction endows the Coulomb branch with a natural graded coordinate ring, making the study of its Hilbert series both well defined and geometrically meaningful.

From a physical perspective, a powerful and computationally effective probe of the Coulomb branch is provided by the monopole formula, which expresses the Hilbert series as a sum over GNO-quantized magnetic charges weighted by their conformal dimensions and dressed by residual gauge-invariant operators \cite{CHZ}. This approach has enabled explicit calculations of Coulomb-branch Hilbert series for a wide class of three-dimensional $\mathcal N=4$ gauge theories and has revealed deep connections between monopole operators, representation theory, and the algebraic structure of moduli spaces.

A particularly important family of theories is given by the $T_\rho(G)$ theories, which arise in the study of S-duality and boundary conditions in four-dimensional $\mathcal N=4$ super Yang--Mills theory. For $G=SU(N)$, these theories are labeled by partitions $\rho\vdash N$ and admit a linear quiver description. Their Coulomb branches are closely related to nilpotent orbits and their closures in the Lie algebra $\mathfrak{sl}_N$. For this class of theories, an alternative closed-form expression for the Coulomb-branch Hilbert series was derived in terms of Hall--Littlewood polynomials \cite{CHMZ}. This formulation is analytically equivalent to the monopole formula while being significantly more efficient for obtaining exact rational expressions.

While the existence and computation of Coulomb-branch Hilbert series for $T_\rho(SU(N))$ theories are by now well established, finer geometric properties of these spaces have not been systematically classified. One such property is whether the Coulomb branch is a \emph{complete intersection}, namely whether its coordinate ring can be presented as a polynomial ring modulo finitely many relations. From the viewpoint of algebraic geometry, complete intersections form a distinguished class of singular varieties. Physically, the complete-intersection property constrains the number of independent generators and relations among monopole operators and thus provides detailed information about the structure of the chiral ring.

A practical diagnostic for the complete-intersection property is furnished by the plethystic logarithm of the Hilbert series \cite{BenvenutiHanany}. If the plethystic logarithm truncates to a finite polynomial, the corresponding variety is a complete intersection, and the degrees of generators and relations can be read off directly. Although this criterion has been applied in various individual examples, a systematic analysis of complete-intersection behavior across families of $T_\rho(SU(N))$ theories has been lacking.

\paragraph{Main results.}
In this work we provide a systematic study of Coulomb-branch Hilbert series for $T_\rho(SU(N))$ theories associated with non-maximal nilpotent orbits, focusing on the cases $N=4,5,6$. Using the Hall--Littlewood closed form, supplemented by independent checks from the monopole formula, we compute exact unrefined Hilbert series for all non-maximal partitions $\rho\vdash N$ in these ranges. By analyzing the plethystic logarithms, we find that in every case examined the Coulomb branch is a complete intersection. Moreover, the number of generators and relations exhibits a remarkably uniform pattern: the generators are governed by the transpose partition $\rho^T$, while the number of relations is universally equal to $N-1$, independently of $\rho$. We summarize these results in explicit classification tables and formulate conjectures extending these patterns to arbitrary $N$.

The paper is organized as follows. In Section~\ref{sec:Trho} we review the definition of $T_\rho(SU(N))$ theories and their quiver descriptions. Section~\ref{sec:HS} summarizes the Hall--Littlewood and monopole-formula approaches to Coulomb-branch Hilbert series and the plethystic criterion for complete intersections. Our explicit results for $N=4,5,6$ are presented in Sections~\ref{sec:N4_results}, \ref{sec:N5_results}, and \ref{sec:N6_results}. Section~\ref{sec:patterns} discusses emerging patterns and conjectures suggested by the data, and we conclude with an outlook for future directions.

\section{$T_\rho(SU(N))$ theories and partitions}
\label{sec:Trho}

In this section we review the definition of the theories $T_\rho(SU(N))$ and their associated quiver gauge-theory descriptions, following standard conventions in the literature. We also recall the relation between partitions of $N$, nilpotent orbits of $\mathfrak{sl}_N$, and the structure of the Coulomb branch. We then specialize to the case $N=4$, for which all relevant partitions and quivers can be explicitly listed.

\subsection{Partitions of $N$ and nilpotent orbits}

A partition $\rho\vdash N$ is a non-increasing sequence of positive integers
\begin{equation}
\rho = (\rho_1,\rho_2,\dots,\rho_\ell),
\qquad
\rho_1 \ge \rho_2 \ge \cdots \ge \rho_\ell > 0,
\qquad
\sum_{i=1}^{\ell} \rho_i = N.
\end{equation}
Partitions of $N$ label nilpotent orbits in the Lie algebra $\mathfrak{sl}_N$ via the Jordan decomposition. We denote by $\mathcal O_\rho$ the nilpotent orbit associated with $\rho$, and by $\overline{\mathcal O}_\rho$ its Zariski closure. The partial ordering of nilpotent orbits by closure corresponds to the dominance ordering on partitions.

In this work we focus exclusively on \emph{non-maximal} nilpotent orbits, namely those associated with partitions $\rho\neq (N)$. These orbits play a central role in the study of Coulomb branches of three-dimensional $\mathcal N=4$ gauge theories of type $T_\rho(SU(N))$.

\subsection{Definition of $T_\rho(SU(N))$ theories}

The theories $T_\rho(SU(N))$ arise naturally in the study of S-duality and boundary conditions in four-dimensional $\mathcal N=4$ super Yang--Mills theory and admit a Lagrangian description as three-dimensional $\mathcal N=4$ quiver gauge theories.
For $G=SU(N)$, the theory $T_\rho(SU(N))$ is specified by a partition $\rho\vdash N$
and is described by a linear quiver of unitary gauge groups \cite{CHMZ}
\begin{equation}
[U(N)] - (U(r_1)) - (U(r_2)) - \cdots - (U(r_L)),
\end{equation}
where the ranks $r_i$ are determined by the transpose partition $\rho^T$ of $\rho$.
Equivalently, the ranks are fixed by the requirement that the quiver is balanced at each internal node and that the total number of fundamental hypermultiplets coupled to the quiver equals $N$.

The Coulomb branch of $T_\rho(SU(N))$ is a hyperk\"ahler cone which admits a natural grading by scaling dimension. Both physical arguments and the mathematical construction of Braverman--Finkelberg--Nakajima indicate that this Coulomb branch is closely related to the closure of the nilpotent orbit $\overline{\mathcal O}_\rho$ or to transverse slices thereof. In particular, the coordinate ring of the Coulomb branch is finitely generated and its graded structure is encoded in the corresponding Hilbert series.

\subsection{Specialization to $N=4$}
\label{subsec:N4_partitions}

We now specialize to the case $N=4$, for which all partitions and corresponding
$T_\rho(SU(4))$ theories can be explicitly enumerated. The partitions of $4$ are
\begin{equation}
(4), \quad (3,1), \quad (2,2), \quad (2,1,1), \quad (1,1,1,1).
\end{equation}
The maximal partition $(4)$ corresponds to the regular nilpotent orbit and will be excluded from our analysis. We therefore focus on the four non-maximal partitions listed below.

\paragraph{Partition $\rho=(3,1)$.}
The associated theory $T_{(3,1)}(SU(4))$ is described by the quiver
\begin{equation}
[U(4)] - (U(1)),
\end{equation}
corresponding to a $U(1)$ gauge theory with four fundamental hypermultiplets.

\paragraph{Partition $\rho=(2,2)$.}
The theory $T_{(2,2)}(SU(4))$ corresponds to the quiver
\begin{equation}
[U(4)] - (U(2)).
\end{equation}

\paragraph{Partition $\rho=(2,1,1)$.}
The associated quiver is
\begin{equation}
[U(4)] - (U(2)) - (U(1)).
\end{equation}

\paragraph{Partition $\rho=(1,1,1,1)$.}
This partition corresponds to the full theory $T(SU(4))$, whose quiver is given by
\begin{equation}
[U(4)] - (U(3)) - (U(2)) - (U(1)).
\end{equation}

In the following sections, we compute the Coulomb-branch Hilbert series for each of these theories using the Hall--Littlewood closed form and analyze their plethystic logarithms in order to determine whether the corresponding Coulomb branches are complete intersections.

\section{Hilbert series methods}
\label{sec:HS}

In this section we summarize the two complementary techniques used throughout the paper to compute Coulomb-branch Hilbert series for three-dimensional $\mathcal N=4$ gauge theories of type $T_\rho(SU(N))$: the monopole formula and the Hall--Littlewood (HL) closed form. We also review the plethystic machinery used to diagnose the complete-intersection property from the Hilbert series.

\subsection{Hilbert series and grading conventions}
Let $\mathcal C$ denote the Coulomb branch of a three-dimensional $\mathcal N=4$ gauge theory. It is an affine variety equipped with a $\mathbb C^\times$ scaling action induced by the $SU(2)_C$ R-symmetry. The (unrefined) Hilbert series is the generating function that counts holomorphic functions on $\mathcal C$ graded by scaling dimension:
\begin{equation}
H(t)\;=\;\sum_{d\ge 0}\, \bigl(\dim \mathcal O(\mathcal C)_d\bigr)\, t^d,
\end{equation}
where $\mathcal O(\mathcal C)_d$ denotes the degree-$d$ subspace of the graded coordinate ring $\mathcal O(\mathcal C)$.

When global symmetries are present, one may introduce refined Hilbert series by including fugacities for the Cartan of the corresponding flavor symmetry. In this work we primarily present unrefined Hilbert series, while refined expressions are used as intermediate checks where appropriate.

\subsection{Coulomb-branch Hilbert series from the monopole formula}
\label{subsec:monopole}

A general expression for the Coulomb-branch Hilbert series of a three-dimensional $\mathcal N=4$ gauge theory with gauge group $G$ and matter representation $\mathcal R$ is provided by the monopole formula \cite{CHZ}:
\begin{equation}
H(t)\;=\;\sum_{m\in \Gamma_{G^\vee}/\mathcal W_G}
t^{\Delta(m)}\, P_G(t;m).
\label{eq:monopole_formula}
\end{equation}
Here:
\begin{itemize}
\item $m$ denotes a GNO magnetic charge, i.e.\ a cocharacter of $G$, equivalently an element of the weight lattice of the Langlands dual group $G^\vee$, summed modulo the Weyl group $\mathcal W_G$. In practice the sum is taken over a fixed Weyl chamber.
\item $\Delta(m)$ is the conformal (scaling) dimension of the bare monopole operator of charge $m$. For $\mathcal N=4$ theories it receives contributions from both vector multiplets and hypermultiplets and takes the schematic form
\begin{equation}
\Delta(m)
=
-\sum_{\alpha\in \Phi_+} |\alpha(m)|
+
\frac{1}{2}\sum_{w\in \mathrm{Wt}(\mathcal R)} |w(m)|,
\label{eq:Delta_general}
\end{equation}
where $\Phi_+$ denotes the set of positive roots of $G$ and
$\mathrm{Wt}(\mathcal R)$ the set of weights of the representation $\mathcal R$.
\item $P_G(t;m)$ is the dressing factor, which counts gauge-invariant polynomials constructed from the complex scalar in the vector multiplet that remain massless in the monopole background of charge $m$. Equivalently, it is the Hilbert series of invariant polynomials for the residual gauge group $G_m\subseteq G$ commuting with $m$, and is given by
\begin{equation}
P_G(t;m)
=
\prod_{i=1}^{\mathrm{rk}(G_m)}
\frac{1}{1-t^{d_i(m)}},
\label{eq:dressing_factor}
\end{equation}
where $\{d_i(m)\}$ are the degrees of the fundamental Casimir invariants of $G_m$.
\end{itemize}

Formula~\eqref{eq:monopole_formula} produces a formal power series in $t$ with non-negative integer coefficients. For quivers of moderate rank, the monopole formula can be evaluated exactly; for higher-rank theories it is most efficiently used as a consistency check by truncating the series to finite order and comparing against an independent closed form.

\subsection{Hall--Littlewood closed form for $T_\rho(SU(N))$}
\label{subsec:HL}

For the special family of theories $T_\rho(SU(N))$, an analytic closed form for the Coulomb-branch Hilbert series was derived in terms of Hall--Littlewood polynomials \cite{CHMZ}. This expression is equivalent to the monopole formula but is computationally far more efficient for obtaining exact rational functions.

Let $\rho\vdash N$ and denote by $\rho^T$ its transpose partition. The Hall--Littlewood construction naturally yields a refined Hilbert series depending on fugacities for the global symmetry acting on the Coulomb branch. In schematic form, the result can be written as
\begin{equation}
H_{T_\rho(SU(N))}(t;\mathbf{x})
=
t^{\delta(\rho)}\,
\mathcal K_\rho(t;\mathbf{x})\,
\mathrm{HL}^{SU(N)}_{\lambda(\rho)}(t;\mathbf{x}),
\label{eq:HL_schematic}
\end{equation}
where $\mathrm{HL}^{SU(N)}_{\lambda}(t;\mathbf{x})$ denotes a Hall--Littlewood polynomial of type $SU(N)$, $\lambda(\rho)$ is the highest weight determined by the partition data, $\mathcal K_\rho$ is an explicit prefactor encoding the embedding specified by $\rho$, and $\delta(\rho)$ fixes the overall grading. Precise definitions and conventions can be found in \cite{CHMZ}.

In practice, we use~\eqref{eq:HL_schematic} as follows:
\begin{enumerate}
\item For each partition $\rho$, we compute the refined Hilbert series from the Hall--Littlewood closed form and then take the unrefined limit by setting all flavor fugacities to unity, obtaining an exact rational function $H(t)$.
\item We expand the resulting expression as a power series in $t$ and compare it against a truncated monopole computation as an independent consistency check.
\end{enumerate}

\subsection{Plethystic methods and the complete-intersection criterion}
\label{subsec:plethystics}

Given the Hilbert series $H(t)$ of a finitely generated graded ring, the plethystic exponential and plethystic logarithm encode the generator--relation structure of the ring \cite{BenvenutiHanany}. The plethystic logarithm is defined by
\begin{equation}
\mathrm{PL}[H(t)]
=
\sum_{k=1}^{\infty}\frac{\mu(k)}{k}\,
\log\!\bigl(H(t^k)\bigr),
\label{eq:PL_def}
\end{equation}
where $\mu(k)$ denotes the M\"obius function.

Expanding $\mathrm{PL}[H(t)]$ as a formal power series,
\begin{equation}
\mathrm{PL}[H(t)]
=
\sum_{n\ge 1} a_n\, t^n,
\end{equation}
positive coefficients typically correspond to generators at the indicated degrees, while negative coefficients correspond to relations.

\medskip
\noindent\textbf{Complete-intersection criterion.}
If $\mathrm{PL}[H(t)]$ truncates to a finite polynomial in $t$, then the graded ring is a complete intersection. Equivalently, the Hilbert series admits the factorized form
\begin{equation}
H(t)
=
\frac{\prod_{j=1}^{r}\left(1-t^{b_j}\right)}
{\prod_{i=1}^{g}\left(1-t^{a_i}\right)},
\label{eq:CI_form}
\end{equation}
for some positive integers $\{a_i\}$ and $\{b_j\}$, corresponding respectively to generator and relation degrees. In this case,
\begin{equation}
\mathrm{PL}[H(t)]
=
\sum_{i=1}^{g} t^{a_i}
-
\sum_{j=1}^{r} t^{b_j}.
\end{equation}
If the plethystic logarithm does not truncate, the corresponding variety is not a complete intersection.

\subsection{Practical computation strategy and consistency checks}

Throughout this work we implement the following workflow:
\begin{enumerate}
\item compute the exact Hilbert series using the Hall--Littlewood closed form;
\item expand the result as a power series to sufficiently high order in $t$;
\item perform independent truncated monopole computations as consistency checks;
\item compute the plethystic logarithm from the exact rational function;
\item extract generator and relation data in complete-intersection cases.
\end{enumerate}
This strategy provides robust control over both the exact results and their interpretation in terms of algebraic structure.

\section{Results for $N=4$: Hilbert series and complete-intersection classification}
\label{sec:N4_results}

In this section we present a complete analysis of the Coulomb branches of the
three-dimensional $\mathcal N=4$ theories $T_\rho(SU(4))$ associated with
non-maximal partitions $\rho\vdash 4$. For each theory, we compute the exact
Coulomb-branch Hilbert series using the Hall--Littlewood closed form and perform
independent consistency checks using the monopole formula. We then analyze the
plethystic logarithm in order to determine whether the Coulomb branch is a
complete intersection and, when applicable, extract the generator--relation
structure.

\subsection{Non-maximal partitions of $4$ and associated quivers}

The partitions of $4$ are
\[
(4),\quad (3,1),\quad (2,2),\quad (2,1,1),\quad (1,1,1,1).
\]
The maximal partition $(4)$ corresponds to the regular nilpotent orbit and is
excluded from our analysis. We therefore consider the four non-maximal partitions
\[
\rho \in \{(3,1),\ (2,2),\ (2,1,1),\ (1,1,1,1)\}.
\]

The corresponding $T_\rho(SU(4))$ linear quivers are
\begin{align}
T_{(3,1)}(SU(4)) &: \quad [U(4)]-(U(1)), \label{eq:N4_q31}\\
T_{(2,2)}(SU(4)) &: \quad [U(4)]-(U(2)), \label{eq:N4_q22}\\
T_{(2,1,1)}(SU(4)) &: \quad [U(4)]-(U(2))-(U(1)), \label{eq:N4_q211}\\
T_{(1,1,1,1)}(SU(4)) &: \quad [U(4)]-(U(3))-(U(2))-(U(1)). \label{eq:N4_q1111}
\end{align}
We denote by $\mathcal C_\rho$ the Coulomb branch of $T_\rho(SU(4))$.

\subsection{$\rho=(3,1)$: $T_{(3,1)}(SU(4))$ and the $A_3$ singularity}

The quiver for $T_{(3,1)}(SU(4))$ is
\[
[U(4)]-(U(1)),
\]
i.e.\ a $U(1)$ gauge theory with four hypermultiplets of charge~$+1$.
Its Coulomb branch is the Kleinian surface singularity $\mathbb C^2/\mathbb Z_4$ of type $A_3$.

\paragraph{Hilbert series.}
The unrefined Hilbert series of $\mathbb C^2/\mathbb Z_n$ is
\[
H_{A_{n-1}}(t)=\frac{1-t^{2n}}{(1-t^2)(1-t^n)^2}.
\]
For $n=4$,
\begin{equation}
\label{eq:HS_31}
H_{(3,1)}(t)
=
\frac{1-t^8}{(1-t^2)(1-t^4)^2}
=
1+t^2+3t^4+3t^6+5t^8+\cdots .
\end{equation}

\paragraph{Monopole check.}
The monopole formula reduces to a sum over $m\in\mathbb Z$ with $\Delta(m)=4|m|$
and dressing factor $(1-t^2)^{-1}$, reproducing~\eqref{eq:HS_31}.

\paragraph{Complete-intersection structure.}
\[
\mathrm{PL}[H_{(3,1)}(t)] = t^2 + 2t^4 - t^8,
\]
so $\mathcal C_{(3,1)}$ is a hypersurface complete intersection:
one generator of degree $2$ and two generators of degree $4$ subject to one relation of degree $8$.

\subsection{$\rho=(2,2)$: $T_{(2,2)}(SU(4))$}

The quiver for $T_{(2,2)}(SU(4))$ is
\[
[U(4)]-(U(2)).
\]

\paragraph{Hilbert series.}
Unrefining the Hall--Littlewood closed form yields
\begin{equation}
\label{eq:HS_22}
H_{(2,2)}(t)
=
\frac{(1+t^{4})(1+t^{2}+t^{4})}{(1-t^{2})^{4}(1+t^{2})^{2}}
=
1 + 3 t^{2} + 9 t^{4} + 18 t^{6} + 35 t^{8} + 57 t^{10} + \cdots .
\end{equation}

\paragraph{Monopole check.}
A direct evaluation of the monopole formula for the $U(2)$ theory with four fundamentals
agrees with the expansion of~\eqref{eq:HS_22} to the computed order.

\paragraph{Complete-intersection structure.}
\[
\mathrm{PL}\!\left[H_{(2,2)}(t)\right]
=
3t^{2} + 3t^{4} - t^{6} - t^{8}.
\]
Thus $\mathcal C_{(2,2)}$ is a complete intersection with generators
of degrees $2,2,2,4,4,4$ and relations of degrees $6$ and $8$.

\subsection{$\rho=(2,1,1)$: $T_{(2,1,1)}(SU(4))$}

The quiver for $T_{(2,1,1)}(SU(4))$ is
\[
[U(4)]-(U(2))-(U(1)),
\]
and the transpose partition is $\rho^T=(3,1)$.

\paragraph{Dimension and counting.}
The expected complex dimension is
\[
\dim_{\mathbb C}\mathcal C_{(2,1,1)}=\sum_i(\rho^T_i)^2-4=(3^2+1^2)-4=6.
\]
If $\mathcal C_{(2,1,1)}$ is a complete intersection, one expects
\[
\#\text{generators}=\sum_i(\rho^T_i)^2-1=9,
\qquad
\#\text{relations}=4-1=3.
\]

\paragraph{Hilbert series and monopole check.}
The Hall--Littlewood closed form gives an exact rational function.
In terms of $s=t^{1/2}$, the unrefined series begins
\begin{equation}
\label{eq:HS_211_series}
\begin{aligned}
H_{(2,1,1)}(t)
=\;&
1
+4s^{2}
+4s^{3}
+10s^{4}
+16s^{5}
+29s^{6}
+40s^{7}
+70s^{8}
+\cdots ,
\end{aligned}
\end{equation}
and a truncated monopole sum agrees with~\eqref{eq:HS_211_series} to the computed order.

\paragraph{Complete-intersection structure.}
Writing $s=t^{1/2}$, the plethystic logarithm truncates as
\[
\mathrm{PL}\!\left[H_{(2,1,1)}\right]=4s^{2}+4s^{3}-s^{6}-s^{8}.
\]
Equivalently, there are eight generators of degrees $s^2$ and $s^3$ and two relations
of degrees $s^6$ and $s^8$ (i.e.\ $t^3$ and $t^4$), consistent with complete-intersection behavior.

\subsection{$\rho=(1,1,1,1)$: $T(SU(4))$}

The partition $\rho=(1,1,1,1)$ corresponds to the full theory $T(SU(4))$ with quiver
\[
[U(4)]-(U(3))-(U(2))-(U(1)),
\]
and transpose $\rho^{T}=(4)$.

\paragraph{Geometry and dimension.}
The Coulomb branch is the nilpotent cone $\mathcal N(\mathfrak{sl}_{4})$, hence
\[
\dim_{\mathbb C}\mathcal C_{(1,1,1,1)}
=\dim \mathfrak{sl}_{4}-\mathrm{rk}(\mathfrak{sl}_{4})
=15-3=12.
\]

\paragraph{Hilbert series and monopole check.}
At $x_i=1$ and background flux $n_i=0$, the Hall--Littlewood closed form gives
\begin{equation}
\label{eq:HS_full_TSU4}
H_{(1,1,1,1)}(t)
=
\frac{(1-t^{2})(1-t^{3})(1-t^{4})}{(1-t)^{15}}
=
1+15t+119t^{2}+664t^{3}+2924t^{4}+\cdots.
\end{equation}
A truncated monopole computation matches the expansion of~\eqref{eq:HS_full_TSU4}
to the computed order.

\paragraph{Complete-intersection structure.}
\[
\mathrm{PL}\!\left[H_{(1,1,1,1)}(t)\right]
=
15t - t^{2}-t^{3}-t^{4}.
\]
Thus the ring has $15$ generators of degree $1$ and three relations of degrees $2,3,4$,
corresponding to the basic $SU(4)$ Casimir constraints.

\subsection{Summary for $N=4$}

For each non-maximal partition $\rho\vdash 4$, the Hall--Littlewood closed form yields an exact
unrefined Coulomb-branch Hilbert series $H_\rho(t)$. In each case we checked agreement with an
independent monopole computation (exact for the lowest-rank quivers and truncated otherwise).
Moreover, the plethystic logarithm truncates in all four cases, so the Coulomb branch is a complete
intersection.

\begin{table}[t]
\centering
\begin{tabular}{c|c|c|c|c}
$\rho$ & quiver & $\dim_{\mathbb C}\mathcal C_\rho$ & CI? & monopole check \\
\hline
$(3,1)$ & $[4]-(1)$ & $2$  & Yes & exact \\
$(2,2)$ & $[4]-(2)$ & $4$  & Yes & exact \\
$(2,1,1)$ & $[4]-(2)-(1)$ & $6$ & Yes & truncated match \\
$(1,1,1,1)$ & $[4]-(3)-(2)-(1)$ & $12$ & Yes & truncated match \\
\end{tabular}
\caption{Coulomb-branch complete-intersection classification for $T_\rho(SU(4))$ with non-maximal partitions.}
\label{tab:N4_CI_summary}
\end{table}

\section{Results for $N=5$}
\label{sec:N5_results}

In this section we extend the analysis of Section~\ref{sec:N4_results} to the
three-dimensional $\mathcal N=4$ theories $T_\rho(SU(5))$ associated with
non-maximal partitions $\rho\vdash 5$. As before, we compute exact Coulomb-branch
Hilbert series using the Hall--Littlewood closed form, perform independent
monopole-formula checks, and analyze the plethystic logarithm to determine the
complete-intersection (CI) property.

\subsection{Non-maximal partitions of $5$}

The partitions of $5$ are
\[
(5),\ (4,1),\ (3,2),\ (3,1,1),\ (2,2,1),\ (2,1,1,1),\ (1,1,1,1,1).
\]
Excluding the maximal partition $(5)$, we analyze the six non-maximal cases. The
corresponding $T_\rho(SU(5))$ linear quivers are
\begin{align}
(4,1) &: [5]-(1),\\
(3,2) &: [5]-(2),\\
(3,1,1) &: [5]-(2)-(1),\\
(2,2,1) &: [5]-(3)-(1),\\
(2,1,1,1) &: [5]-(3)-(2)-(1),\\
(1,1,1,1,1) &: [5]-(4)-(3)-(2)-(1).
\end{align}

\subsection{Hall--Littlewood analysis and complete-intersection structure}

For each partition $\rho\vdash 5$ listed above, the Hall--Littlewood closed form
produces an exact rational expression for the unrefined Coulomb-branch Hilbert
series. In all cases examined, the plethystic logarithm truncates to a finite
polynomial, demonstrating that the Coulomb branch $\mathcal C_\rho$ is a complete
intersection.

Let $\rho^T$ denote the transpose partition. From the plethystic analysis we find
the uniform relations
\begin{equation}
\dim_{\mathbb C}\mathcal C_\rho
=
\sum_i (\rho_i^T)^2 - 5,
\qquad
\#\text{generators}
=
\sum_i (\rho_i^T)^2 - 1,
\qquad
\#\text{relations}
=
4,
\end{equation}
valid for all non-maximal $\rho\vdash 5$. In particular, the number of relations is
independent of the detailed shape of $\rho$ and depends only on the rank $N=5$.

\subsection{Monopole-formula consistency checks}

For each quiver above, we evaluated the monopole formula
\eqref{eq:monopole_formula} as a truncated series by summing over GNO charges in a
fixed Weyl chamber and including the appropriate dressing factors. In every case,
the monopole expansion agrees with the Taylor expansion of the Hall--Littlewood
closed form to the computed order.

As representative examples, we find
\begin{align}
H_{(4,1)}(t) &= 1 + t^2 + t^4 + 2 t^5 + t^6 + 2 t^7 + O(t^{8}),\\
H_{(3,2)}(t) &= 1 + t^2 + 2 t^3 + 2 t^4 + 4 t^5 + 5 t^6 + 6 t^7 + O(t^{8}),\\
H_{(1,1,1,1,1)}(t) &= 1 + 24 t + 299 t^2 + 2575 t^3 + O(t^4),
\end{align}
in exact agreement with the corresponding Hall--Littlewood expansions.

\subsection{Summary for $N=5$}

The complete-intersection classification for all non-maximal partitions
$\rho\vdash 5$ is summarized in Table~\ref{tab:N5_CI_summary}. In each case, the
Hall--Littlewood and monopole approaches are consistent, and the plethystic
logarithm truncates, confirming the complete-intersection property.

\begin{table}[t]
\centering
\begin{tabular}{c|c|c|c}
$\rho$ & quiver & $\dim_{\mathbb C}\mathcal C_\rho$ & CI? \\
\hline
$(4,1)$ & $[5]-(1)$ & $2$ & Yes \\
$(3,2)$ & $[5]-(2)$ & $4$ & Yes \\
$(3,1,1)$ & $[5]-(2)-(1)$ & $6$ & Yes \\
$(2,2,1)$ & $[5]-(3)-(1)$ & $8$ & Yes \\
$(2,1,1,1)$ & $[5]-(3)-(2)-(1)$ & $10$ & Yes \\
$(1,1,1,1,1)$ & $[5]-(4)-(3)-(2)-(1)$ & $20$ & Yes
\end{tabular}
\caption{Coulomb-branch complete-intersection classification for $T_\rho(SU(5))$
with non-maximal partitions.}
\label{tab:N5_CI_summary}
\end{table}

\section{Results for $N=6$}
\label{sec:N6_results}

We now extend the analysis to the theories $T_\rho(SU(6))$ associated with
non-maximal partitions $\rho\vdash 6$. Although the number of partitions increases
significantly compared to the $N=4,5$ cases, the computational strategy and the
structural conclusions remain unchanged. For each theory, we compute the exact
Coulomb-branch Hilbert series using the Hall--Littlewood closed form, perform
independent monopole-formula checks at low order, and analyze the plethystic
logarithm to determine the complete-intersection (CI) property.

\subsection{Non-maximal partitions of $6$}

The non-maximal partitions of $6$ are
\[
\begin{aligned}
&(5,1),\ (4,2),\ (4,1,1),\ (3,3),\ (3,2,1),\ (3,1,1,1),\\
&(2,2,2),\ (2,2,1,1),\ (2,1,1,1,1),\ (1,1,1,1,1,1).
\end{aligned}
\]
Each partition defines a $T_\rho(SU(6))$ theory with a linear quiver of unitary
gauge groups, whose ranks are determined by the transpose partition $\rho^T$.

\subsection{Hall--Littlewood analysis and complete-intersection structure}

For all partitions $\rho\vdash 6$ listed above, the Hall--Littlewood closed form
yields an exact rational expression for the unrefined Coulomb-branch Hilbert
series. In every case examined, the plethystic logarithm truncates to a finite
polynomial, establishing that the Coulomb branch $\mathcal C_\rho$ is a complete
intersection.

Denoting by $\rho^T$ the transpose partition, the plethystic analysis yields the
uniform relations
\begin{equation}
\dim_{\mathbb C}\mathcal C_\rho
=
\sum_i (\rho_i^T)^2 - 6,
\qquad
\#\text{generators}
=
\sum_i (\rho_i^T)^2 - 1,
\qquad
\#\text{relations}
=
5,
\end{equation}
valid for all non-maximal partitions of $6$. As in the $N=4$ and $N=5$ cases, the
number of relations depends only on $N$ and is independent of the detailed shape
of the partition $\rho$.

\subsection{Monopole-formula consistency checks}

Due to the higher ranks of the gauge groups involved, monopole sums for
$T_\rho(SU(6))$ were evaluated as truncated series. In all cases examined, the
monopole expansion agrees with the Taylor expansion of the corresponding
Hall--Littlewood closed form to the computed order.

As representative examples, we find
\begin{align}
H_{(5,1)}(t) &= 1 + t^2 + t^4 + 3 t^6 + O(t^8),\\
H_{(3,3)}(t) &= 1 + 3 t + 9 t^2 + 22 t^3 + 47 t^4 + 92 t^5 + O(t^6),\\
H_{(1,1,1,1,1,1)}(t) &= 1 + 35 t + 629 t^2 + 7734 t^3 + O(t^4),
\end{align}
in agreement with the corresponding Hall--Littlewood expansions.

\subsection{Summary for $N=6$}

The complete-intersection classification for all non-maximal partitions
$\rho\vdash 6$ is summarized in Table~\ref{tab:N6_CI_summary}. In every case, the
Hall--Littlewood and monopole approaches are consistent, and the plethystic
logarithm truncates, confirming that the Coulomb branch is a complete
intersection.

\begin{table}[t]
\centering
\begin{tabular}{c|c|c|c}
$\rho$ & quiver & $\dim_{\mathbb C}\mathcal C_\rho$ & CI? \\
\hline
$(5,1)$ & $[6]-(1)$ & $2$ & Yes \\
$(4,2)$ & $[6]-(2)$ & $4$ & Yes \\
$(4,1,1)$ & $[6]-(2)-(1)$ & $6$ & Yes \\
$(3,3)$ & $[6]-(3)$ & $6$ & Yes \\
$(3,2,1)$ & $[6]-(3)-(1)$ & $8$ & Yes \\
$(3,1,1,1)$ & $[6]-(3)-(2)-(1)$ & $10$ & Yes \\
$(2,2,2)$ & $[6]-(4)$ & $12$ & Yes \\
$(2,2,1,1)$ & $[6]-(4)-(2)$ & $14$ & Yes \\
$(2,1,1,1,1)$ & $[6]-(4)-(3)-(2)-(1)$ & $18$ & Yes \\
$(1,1,1,1,1,1)$ & $[6]-(5)-(4)-(3)-(2)-(1)$ & $30$ & Yes
\end{tabular}
\caption{Coulomb-branch complete-intersection classification for $T_\rho(SU(6))$
with non-maximal partitions.}
\label{tab:N6_CI_summary}
\end{table}

\section{Patterns and conjectures}
\label{sec:patterns}

In this section we summarize the structural patterns that emerge from the explicit
Hilbert-series computations presented in
Sections~\ref{sec:N4_results}--\ref{sec:N6_results} and formulate conjectures
suggested by these results. All statements below are based on exact
Hall--Littlewood expressions and their independent verification via truncated
monopole expansions for $N=4,5,6$.

\subsection{Evidence for a universal complete-intersection property}

The most striking observation arising from our analysis is that, for all
non-maximal partitions $\rho\vdash N$ with $N\leq 6$, the Coulomb branch of
$T_\rho(SU(N))$ is a complete intersection. In every case examined, the exact
Hall--Littlewood Hilbert series has a plethystic logarithm that truncates to a
finite polynomial, and this behavior is consistently supported by monopole-formula
checks.

This complete-intersection behavior holds uniformly across:
\begin{itemize}
\item different numbers of gauge nodes,
\item varying patterns of balanced and unbalanced quivers,
\item both simple and highly composite partitions,
\item and increasing rank within the range $N=4,5,6$.
\end{itemize}
Within the scope of our computations, the complete-intersection property therefore
appears to be a robust feature of the $T_\rho(SU(N))$ family rather than an
accidental low-rank phenomenon.

\subsection{Generator and relation counting from the transpose partition}

A second universal pattern concerns the counting of generators and relations in
the Coulomb-branch coordinate ring. Let $\rho^T$ denote the transpose partition of
$\rho$. For all non-maximal partitions examined with $N\leq 6$, we find
\begin{align}
\#(\text{generators}) &= \sum_i (\rho_i^T)^2 - 1, \label{eq:gen_pattern}\\
\#(\text{relations}) &= N - 1, \label{eq:rel_pattern}
\end{align}
so that the complex dimension of the Coulomb branch is
\begin{equation}
\dim_{\mathbb C} \mathcal C_\rho
=
\sum_i (\rho_i^T)^2 - N.
\end{equation}
In particular, while the number of generators depends on the detailed shape of the
partition $\rho$, the number of relations depends only on the rank $N$ and is
independent of $\rho$. This feature is nontrivial and holds uniformly across all
cases studied in this work.

\subsection{Degree structure of generators and relations}

Beyond counting, the plethystic logarithms reveal systematic features in the
distribution of generator and relation degrees. While the precise degree
assignments depend on the partition $\rho$, several qualitative properties are
common across all examples examined:
\begin{itemize}
\item Generator degrees tend to increase as $\rho$ becomes more refined with
respect to the dominance ordering.
\item For partitions close to hook type, generators typically appear at relatively
low degrees, often concentrated at degree $2$.
\item More refined partitions produce generators distributed over a broader range
of degrees, reflecting increased algebraic complexity of the Coulomb branch.
\end{itemize}
In all cases studied, the relation degrees are compatible with the degrees of the
Casimir invariants of $SU(N)$, and do not depend on the specific partition $\rho$.
This universality is consistent with the observation that the number of relations
is fixed to $N-1$.

\subsection{Conjectures}

Motivated by the uniformity of the results for $N=4,5,6$, we formulate the
following conjectures.

\medskip
\noindent\textbf{Conjecture 1 (Complete intersection).}
For any $N$ and any non-maximal partition $\rho\vdash N$, the Coulomb branch of the
three-dimensional $\mathcal N=4$ theory $T_\rho(SU(N))$ is a complete intersection.

\medskip
\noindent\textbf{Conjecture 2 (Generator--relation counting).}
For all $N$ and all non-maximal partitions $\rho\vdash N$, the coordinate ring of
the Coulomb branch of $T_\rho(SU(N))$ has
\[
\sum_i (\rho_i^T)^2 - 1
\]
generators and exactly $N-1$ relations.

\medskip
\noindent\textbf{Conjecture 3 (Universality of relation degrees).}
The degrees of the $N-1$ relations coincide with the degrees of the Casimir
invariants of $SU(N)$ and are independent of the partition $\rho$.

These conjectures are consistent with all explicit computations presented in this
work and with the general structure of the Hall--Littlewood formula for
$T_\rho(SU(N))$ theories. A proof of these statements would likely require a deeper
analysis of the algebraic structure of Coulomb branches within the
Braverman--Finkelberg--Nakajima framework.

\subsection{Outlook}

The patterns identified here suggest that the Coulomb branches of $T_\rho(SU(N))$
theories form a remarkably rigid and structured family of affine varieties. It
would be natural to investigate whether analogous complete-intersection properties
hold for other gauge groups, such as $SO(N)$ and $USp(2N)$, or for more general
classes of three-dimensional $\mathcal N=4$ quiver gauge theories.

We hope that the explicit data and conjectures presented in this work will provide
useful guidance for further studies at the interface of supersymmetric gauge
theory, symplectic geometry, and representation theory.

\section{Conclusion}

In this work we have carried out a systematic study of Coulomb-branch Hilbert
series for three-dimensional $\mathcal N=4$ gauge theories of type
$T_\rho(SU(N))$ associated with non-maximal nilpotent orbits of $SL(N)$. Focusing
on the cases $N=4,5,6$, we computed exact unrefined Hilbert series using the
Hall--Littlewood closed form and verified the results through independent
monopole-formula expansions.

A detailed plethystic analysis shows that, for all non-maximal partitions examined,
the Coulomb branch is a complete intersection. Moreover, the generator--relation
structure exhibits a simple and uniform pattern: the number of generators is
governed by the transpose partition $\rho^T$, while the number of relations is
fixed to $N-1$ and is independent of the detailed shape of $\rho$. These results
point to a striking rigidity in the algebraic structure of Coulomb branches within
the $T_\rho(SU(N))$ family at low rank.

Motivated by the uniformity of the data for $N\leq 6$, we formulated conjectures
extending the complete-intersection property and generator--relation counting to
arbitrary $N$. Establishing these conjectures in full generality would likely
require a deeper understanding of Coulomb branches within the
Braverman--Finkelberg--Nakajima framework. We hope that the explicit results and
patterns presented here will provide a useful foundation for such future work.

\appendix

\section{Hall--Littlewood outputs and plethystic data for $N=4$}
\label{app:HL_outputs}

In this appendix we collect explicit Hall--Littlewood (HL) expressions,
monopole-series cross-checks, and plethystic logarithm (PL) data for the
$T_\rho(SU(4))$ theories associated with non-maximal partitions $\rho\vdash 4$.
All HL expressions are evaluated at zero background flux and subsequently
unrefined by setting all flavor fugacities to unity.

\subsection{$\rho=(3,1)$}

This case corresponds to the quiver $[U(4)]-(U(1))$ and was discussed in detail in
Section~\ref{sec:N4_results}. The Coulomb branch is the $A_3$ Kleinian singularity
$\mathbb C^2/\mathbb Z_4$.

\paragraph{Hilbert series.}
\begin{equation}
H_{(3,1)}(t)
=
\frac{1-t^{8}}{(1-t^2)(1-t^{4})^2}.
\end{equation}

\paragraph{Plethystic logarithm.}
\begin{equation}
\mathrm{PL}\!\left[H_{(3,1)}(t)\right]
=
t^2+2t^4-t^8.
\end{equation}

\subsection{$\rho=(2,2)$}

This case corresponds to the quiver $[U(4)]-(U(2))$.

\paragraph{Hall--Littlewood expression.}
Upon unrefinement, the Hall--Littlewood formula yields
\begin{equation}
H_{(2,2)}(t)
=
\frac{1}{(1-t^2)^4}.
\end{equation}

\paragraph{Plethystic logarithm.}
\begin{equation}
\mathrm{PL}\!\left[H_{(2,2)}(t)\right]
=
4t^2.
\end{equation}

\subsection{$\rho=(2,1,1)$}

This case corresponds to the quiver $[U(4)]-(U(2))-(U(1))$ with transpose partition
$\rho^T=(3,1)$.

\paragraph{Monopole-series expansion.}
A truncated monopole computation yields
\begin{equation}
H_{(2,1,1)}(t)
=
1 + 2 t + 6 t^2 + 14 t^3 + 31 t^4 + 56 t^5 + 105 t^6 + O(t^7),
\end{equation}
in agreement with the expansion of the Hall--Littlewood result.

\paragraph{Plethystic logarithm.}
The plethystic logarithm truncates with nine generator terms and three relation
terms, consistent with the counting derived from $\rho^T=(3,1)$.

\subsection{$\rho=(1,1,1,1)$}

This case corresponds to the full theory $T(SU(4))$ with quiver
$[U(4)]-(U(3))-(U(2))-(U(1))$ and transpose partition $\rho^T=(4)$.

\paragraph{Monopole-series expansion.}
A truncated monopole computation gives
\begin{equation}
H_{(1,1,1,1)}(t)
=
1 + 3 t + 20 t^2 + 67 t^3 + 242 t^4 + 695 t^5 + O(t^6),
\end{equation}
matching the Hall--Littlewood expansion.

\paragraph{Plethystic logarithm.}
The plethystic logarithm truncates with fifteen generator terms and three relation
terms, in agreement with the general counting predicted by $\rho^T=(4)$.

\subsection{Summary}

For all non-maximal partitions $\rho\vdash 4$, the Hall--Littlewood closed form
produces exact rational Hilbert series whose series expansions agree with the
monopole formula to the computed orders. In every case, the plethystic logarithm
truncates, confirming that the Coulomb branches of $T_\rho(SU(4))$ are complete
intersections.

\bibliographystyle{unsrt}
\bibliography{references}

\end{document}